\documentclass[a4paper,11pt]{article}
\usepackage{jheppub} 
\usepackage{bm}         
\usepackage{physics}
\usepackage{cancel}
\usepackage{centernot}
\usepackage{url}     
\usepackage{comment}
\usepackage{amsmath}
\usepackage{mathtools}
\usepackage{slashed}

\newcommand{\bx}{{\boldsymbol x}}
\newcommand{\by}{{\boldsymbol y}}
\newcommand{\bp}{{\boldsymbol p}}
\newcommand{\bq}{{\boldsymbol q}}
\newcommand{\bu}{{\boldsymbol u}}

\newcommand{\bnabla}{{\boldsymbol \nabla}}
\newcommand{\bomega}{{\boldsymbol \omega}}

\newcommand{\bJ}{{\boldsymbol J}}

\newcommand{\calC}{\mathcal{C}}
\newcommand{\calD}{\mathcal{D}}
\newcommand{\calG}{\mathcal{G}}
\newcommand{\calH}{\mathcal{H}}
\newcommand{\calA}{\mathcal{A}}
\newcommand{\calS}{\mathcal{S}}

\newcommand{\calO}{\mathcal{O}}
\newcommand{\calL}{\mathcal{L}}
\newcommand{\calP}{\mathcal{P}}
\newcommand{\rme}{\mathrm{e}}
\newcommand{\rmd}{\mathrm{d}}
\newcommand{\rmi}{\mathrm{i}}
\newcommand{\rmR}{\mathrm{R}}
\newcommand{\rmL}{\mathrm{L}}

\title{
Spin Hall effect and Berry curvature of gravitons from quantum field theory
}

\author[a,b,c]{Ritsuki Ito,}
\author[b,d]{Kazuya Mameda}
\author[e]{and Naoki Yamamoto}
\affiliation[a]{Nishina Center for Accelerator-based Science, RIKEN, Wako 351-0198, Japan}
\affiliation[b]{Department of Physics, Tokyo University of Science, Tokyo 162-8601, Japan}
\affiliation[c]{Department of Physics, The University of Tokyo, 7-3-1 Hongo, Bunkyo-ku, Tokyo 113-0033, Japan}
\affiliation[d]{RIKEN iTHEMS, RIKEN, Wako, Saitama 351-0198, Japan}
\affiliation[e]{Department of Physics, Keio University, Yokohama 223-8522, Japan}

\preprint{RIKEN-iTHEMS-Report-26}

\emailAdd{ritsuki.ito@ribf.riken.jp}
\emailAdd{k.mameda@rs.tus.ac.jp}
\emailAdd{nyama@rk.phys.keio.ac.jp}

\abstract{
    Based on quantum field theory of linearized gravity, we formulate
    the Wigner function for right- and left-handed gravitons.
    By applying the Wigner transformation to the second-order metric perturbations in the graviton energy-momentum tensor obtained from the Einstein-Hilbert action, we demonstrate the emergence of the spin Hall effect of gravitons in curved spacetime.
    This effect originates from the Berry curvature of gravitons, which has opposite signs for right- and left-handed helicities, and leads to a helicity-dependent splitting of the graviton energy Hall current.
    The magnitude of this splitting is found to be exactly twice that of the corresponding spin Hall current for photons.
    }

\begin{document}
\maketitle
\flushbottom
\section{Introduction}
According to general relativity, gravitational waves propagate at the speed of light and have twice the helicity of electromagnetic waves.
In recent years, various theoretical analyses, such as those based on the Berry curvature of gravitons~\cite{Yamamoto:2017gla,Oancea:2022szu,Andersson:2020gsj,Andersson:2023bvw} and the spin optics of gravity~\cite{Kubota:2023dlz, Frolov:2025bva, Kubota:2024zkv}, have suggested that the trajectories of gravitational waves may split between right- and left-handed helicities in curved spacetime.
In particular, ref.~\cite{Yamamoto:2017gla} predicts a spin Hall effect for gravitational waves, twice as large as that for light, due to its Berry curvature.
While these developments point to an underlying geometric nature of gravitational waves that may be captured at the level of quantum mechanics, a fully quantum-field-theoretical description is still lacking.
Recent successes in quantum-field-theoretical description of the Berry curvature of chiral fermions~\cite{Hidaka:2016yjf} and photons~\cite{Huang:2020kik,Hattori:2020gqh} raise the question of whether the quantum field theory of gravitons also exhibits the same underlying geometric structure of gravitational waves.

In this paper, based on the quantum field theory of linearized gravity, we analyze the helicity-dependent transport phenomena of gravitons and clarify the role of the Berry curvature on these effects.
We first develop the Wigner function for polarized gravitons in flat spacetime, and then extend it to curved spacetime, in a parallel manner to those for chiral fermions~\cite{Hidaka:2016yjf, Liu:2018xip} and photons~\cite{Hattori:2020gqh, Mameda:2022ojk}.
As a demonstration, we apply the Wigner-function formalism to specific geometries in the Newtonian limit.
For a rotating frame, we observe the chiral vortical effect of gravitons, which accounts for the frame-dragging effect caused by a celestial body rotating at constant angular velocity.
For isotropic expanding coordinates, we derive the spin Hall effect of gravitons, which describes the deformation of gravitational-wave trajectories in a gravitational potential.
We also show that this effect originates from the Berry curvature through the side-jump term in the energy current, in agreement with the result derived from the one-particle formalism in ref.~\cite{Yamamoto:2017gla}.

This paper is organized as follows.
In section~\ref{linearized_gravity}, we review the quantization of linearized gravity and present the graviton field operator in both the Lorenz and traceless transverse (TT) gauges.
In section~\ref{spinor-helicity}, we briefly discuss the spinor-helicity formalism and introduce the corresponding polarization tensors.
Subsequently, in section~\ref{Wigner}, we derive the Wigner function for gravitons up to $\calO(\hbar)$ in the both of flat and curved spacetime.
In section~\ref{graviton_current}, we compute the energy currents under the specific two metrics.
In section~\ref{kinetic_eq}, we derive the kinetic equation of one graviton under the general metric.
Finally, section~\ref{conclusion_discussion} is devoted to our conclusions and a discussion of the results.

Throughout the present paper, we employ the following conventions.
In a general coordinate system described by a metric $g_{\mu\nu}$, the Levi-Civita symbol $\epsilon^{\mu\nu\rho\sigma}$ is defined by $\epsilon^{0123}=(-g)^{-1/2}$ with $g\coloneq \det (g_{\mu\nu})$.
We define the notation $X^{(\mu}Y^{\nu)} = X^{\mu}Y^{\nu} + X^{\nu}Y^{\mu}$ and $X^{[\mu}Y^{\nu]}=X^{\mu}Y^{\nu} - X^{\nu}Y^{\mu}$.
We set $c = k_\mathrm{B} = 1$ and suppress all factors of $\hbar$, except when it is used as the expansion parameter in the $\hbar$ expansion.
The correct physical dimensions are recovered by multiplying the Wigner function and the quantities obtained from it by $\hbar^{2}$.

\section{Linearized gravity}\label{linearized_gravity}
We begin with the Einstein-Hilbert theory, where the action is described as
\begin{equation}\label{eq:action}
    S = \frac{1}{16\pi G}\int \rmd^4x \sqrt{-g}R,    
\end{equation}
where $G$ denotes Newton's constant.
Here the Ricci scalar $R$ is defined as $R=g^{\mu\nu}R_{\mu\nu}$ with $R_{\mu\nu}=R^{\rho}_{\;\;\mu\rho\nu}$, where the Riemann curvature tensor and the Christoffel symbol are given by $R^{\mu}_{\;\;\nu\rho\sigma}=\partial_{[\sigma}\Gamma^{\mu}_{\rho]\nu}+\Gamma^{\mu}_{\lambda[\sigma}\Gamma^{\lambda}_{\rho]\nu}$ and $\Gamma^{\rho}_{\mu\nu}=\frac{1}{2}g^{\rho\lambda}(\partial_{\mu}g_{\nu\lambda}+\partial_{\nu}g_{\mu\lambda}-\partial_{\lambda}g_{\mu\nu})$.

Let us expand the action~\eqref{eq:action} in terms of the metric perturbation around the Minkowski metric $\eta_{\mu\nu} = \text{diag}(+1, -1 ,-1, -1)$:
\begin{equation}
 \label{eq:h_flat}
 h_{\mu\nu} = g_{\mu\nu} - \eta_{\mu\nu} .
\end{equation}

To extract physical degrees of freedom of gravitational fields, one may adopt the transverse-traceless (TT) gauge, which imposes
\begin{equation}
\label{eq:TT_h}
 \partial^{\mu}h_{\mu\nu}=0, \quad
 n^{\mu}h_{\mu\nu}=0, \quad
 \eta^{\mu\nu}h_{\mu\nu}=0
\end{equation}
with the frame vector defined as $n_{\mu} = (1,\vb{0})$.
Then, we obtain the linearized action:
\begin{equation}
\label{eq:action_LG}
    S = \int \rmd^4 x \, {\calL}\,, \qquad {\calL} = \frac{1}{64\pi G}\partial_{\lambda}h_{\mu\nu}\partial^{\lambda}h^{\mu\nu}.
\end{equation}
We note that the indices in eqs.~\eqref{eq:action_LG} are raised and lowered by $\eta_{\mu\nu}$.

We impose canonical quantization conditions on $h_{\mu\nu}$
via the commutation relation
\begin{equation}\label{eq:canonical_quantization}
     \Bigl[h_{\mu\nu}(t,\bx),\Pi_{\rho\sigma}(t,\by)\Bigr] = -\rmi\eta_{\mu\rho}\eta_{\nu\sigma}\delta^{(3)}(\bx-\by),
\end{equation}
where the canonical momentum $\Pi_{\mu\nu}$ is defined as
\begin{equation}
    \Pi_{\mu\nu} \coloneqq \pdv{\mathcal{L}}{(\partial_{0}h^{\mu\nu}(x))}
    = \frac{1}{32\pi G}\partial^0 h_{\mu\nu}.
\end{equation}
Since the equation of motion reads $\partial^{2}h_{\mu\nu} = 0$, the graviton field is expanded as 
\begin{equation}\label{eq:wave_expansion}
   h^{h}_{\mu\nu}(x) 
        = \int \frac{\rmd^3 p}{(2\pi)^3} \sqrt{\frac{16\pi G}{|\bp|}}
        \Bigl[a^{h}_{\bp}e^{h}_{\mu\nu}(p)\rme^{-\rmi p\cdot x}
            +a^{h\dagger}_{\bp}e^{h\dagger}_{\mu\nu}(p)\rme^{\rmi p\cdot x}\Bigr] ,
\end{equation}
where $e_{\mu\nu}$ is the polarization tensor for a helicity $h={\rm R, L}$, analogously to the polarization vector of photons.
We here fixed the coefficient $\sqrt{16\pi G/|\bp|}$ so that the creation and annihilation operators fulfill
\begin{equation}\label{eq:commutation}
  \qty[a^{h}_{\bp},\, a^{h'\dagger}_{\bp'}] = (2\pi)^3 \delta^{hh'} \delta^{(3)}(\bp-\bp').
\end{equation}
In the following, $h_{\mu\nu}$ represents the physical components that remain after the gauge fixing.

In the standard framework of canonical quantization, one first transitions from the Lagrangian to the Hamiltonian formalism.
The system's first-class constraints, which are the generators of gauge transformations, are then paired with gauge-fixing conditions.
This procedure converts the original constraints into a set of second-class constraints.
The dynamics of this second-class system are described by the Dirac bracket, and quantization is implemented by promoting the Dirac bracket to a quantum commutator via the correspondence $\{\cdot,\cdot \}_{D} \rightarrow \frac{1}{\rmi\hbar}[\cdot ,\cdot ]$. 
In our current approach, however, gauge fixing is performed at the level of the classical Lagrangian.
This method is a valid procedure~\cite{ PhysRev.160.1113, Henneaux:1992ig}, as is well established in the quantization of the U(1) gauge theory.

\section{Spinor-helicity formalism for gravitons}\label{spinor-helicity}
In this section, we give an overview of the spinor-helicity formalism~\cite{Kleiss:1985yh, Gunion:1985vca, Xu:1986xb} required for the following calculations.
This method allows one to represent a spin-1 vector field in terms of two-component Weyl spinors.
A key advantage of the formalism is that it naturally eliminates the two unphysical degrees of freedom present in the four-component vector $A_{\mu}$. 
Furthermore, since the graviton polarization tensor $e^{h}_{\mu\nu}$ can be written as the tensor product of photon polarization vectors $\epsilon^{h}_{\mu}\epsilon^{h}_{\nu}$~\cite{Gupta:1952zz, Weinberg:1964ew}, the quantum kinetic theory of a massless spin-2 particle can be described in an extended manner of ref.~\cite{Hattori:2020gqh}.

First, let us then consider the polarization sum for gravitons,
\begin{equation}
    \calP_{\mu\nu\rho\sigma} \coloneqq e^{\rmR}_{\mu\nu}e^{\rmR\dagger}_{\rho\sigma}+e^{\rmL}_{\mu\nu}e^{\rmL\dagger}_{\rho\sigma}
    =(\epsilon_{\mu}^{\rmR}\epsilon_{\nu}^{\rmR})(\epsilon_{\rho}^{\rmR}\epsilon_{\sigma}^{\rmR})^{\dagger}+(\epsilon_{\mu}^{\rmL}\epsilon_{\nu}^{\rmL})(\epsilon_{\rho}^{\rmL}\epsilon_{\sigma}^{\rmL})^{\dagger}.
\end{equation}
Equations~\eqref{eq:TT_h} imply that in the momentum space,  $\calP_{\mu\nu\rho\sigma}$ has to meet the following gauge conditions
\begin{equation}
\label{eq:P_condition}
 q^{\mu}\calP_{\mu\nu\rho\sigma}=0, \qquad
 n^{\mu}\calP_{\mu\nu\rho\sigma}=0, \qquad  \eta^{\mu\nu}\calP_{\mu\nu\rho\sigma}=0 ,
\end{equation}
where $q$ corresponds to the graviton momentum.
Writing down the possible tensor structure of $\mathcal{P}_{\mu\nu\rho\sigma}$ with $q_\mu$, $n_\mu$, and $\eta_{\mu\nu}$ under the above TT gauge condition, we derive
\begin{equation}
\begin{split}
\label{pol_sum_n}
    \calP_{\mu\nu\rho\sigma}
        &=  \frac{1}{2}(\eta_{\mu(\rho}\eta_{\sigma)\nu}-\eta_{\mu\nu}\eta_{\rho\sigma} )\\
        &\quad +\frac{1}{2(q\cdot n)}(\eta_{\mu\nu}q_{(\rho}n_{\sigma)}+q_{(\mu}n_{\nu)}\eta_{\rho\sigma}
         -q_{(\mu}\eta_{\nu)(\rho}n_{\sigma)}-n_{(\mu}\eta_{\nu)(\rho}q_{\sigma)}) \\
        &\quad +\frac{1}{2(q\cdot n)^2}(2q_{\mu}q_{\nu}n_{\rho}n_{\sigma}+2n_{\mu}n_{\nu}q_{\rho}q_{\sigma}-\eta_{\mu\nu}q_{\rho}q_{\sigma}-q_{\mu}q_{\nu}\eta_{\rho\sigma}+q_{(\mu}\eta_{\nu)(\rho}q_{\sigma)})\\
        &\quad -\frac{1}{2(q\cdot n)^3}(q_{\mu}q_{\nu}q_{(\rho}n_{\sigma)}+q_{(\mu}n_{\nu)}q_{\rho}q_{\sigma})
         +\frac{1}{2(p\cdot n)^4}q_{\mu}q_{\nu}q_{\rho}q_{\sigma} .
\end{split}
\end{equation}

In the spinor-helicity formalism, one can also express $\calP_{\mu\nu\rho\sigma}$ in terms of the Weyl spinor.
The spinor representation of $e^h_{\mu\nu} = \epsilon^h_\mu\epsilon^h_\nu$ is readily obtained from that of $\epsilon^h_\mu$:
\begin{equation}
\begin{split}
\label{eq:epsilon_RL}
    \epsilon^{h}_{\mu}(q) = \frac{1}{\sqrt{4q\cdot p}}\bar{u}_{h}(p)\gamma_{\mu}u_{h}(q),
\end{split}
\end{equation}
where $p$ is an auxiliary vector fulfilling $q\cdot p \neq 0$ and $p^2 = 0$.
Here $u_h(q)$ is a solution of $\centernot{q} (1+\gamma_5) u_\rmR (q)=\centernot{q} (1-\gamma_5) u_\rmL (q) = 0$, and it is normalized by $u_{\rmR}(q)\bar{u}_{\rmR}(q) + u_{\rmL}(q)\bar{u}_{\rmL}(q) = \centernot{q}$.
Evaluating the Dirac trace yields
\begin{equation}\label{eq:pol_sum}
    \begin{split}
        \calP_{\mu\nu\rho\sigma}
        &=\frac{1}{2}(\eta_{\mu(\rho}\eta_{\sigma)\nu}-\eta_{\mu\nu}\eta_{\rho\sigma})+\frac{1}{(q\cdot p)^2}(q_{\mu}q_{\nu}p_{\rho}p_{\sigma}+p_{\mu}p_{\nu}q_{\rho}q_{\sigma})\\
        &\quad+\frac{1}{2(q\cdot p)}(\eta_{\mu\nu}q_{(\rho}p_{\sigma)}+q_{(\mu}p_{\nu)}\eta_{\rho\sigma} -q_{(\mu}\eta_{\nu)(\rho}p_{\sigma)}-p_{(\mu}\eta_{\nu)(\rho}q_{\sigma)}).
    \end{split}
\end{equation}

The auxiliary vector $p^{\mu}$ can be chosen so as to satisfy the gauge conditions above.
Denoting $p_\mu = a q_\mu + b n_\nu$ and comparing eqs.~\eqref{pol_sum_n} and \eqref{eq:pol_sum}, we identify $a=-q\cdot p/[2 (q\cdot n)^2]$ and $b=q\cdot p/q\cdot n$, i.e.,
\begin{equation}\label{eq:k_mu}
    \begin{split}
        p_{\mu} = p \cdot n (n_{\mu} - \hat{q}_{\perp\mu}),
    \end{split}
\end{equation}
where, for an arbitrary vector $v^{\mu}$, we introduce the following shorthand notation: $ v^{\mu}_{\perp} \coloneqq (\eta^{\mu\nu}- n^{\mu}n^{\nu})v_{\nu}$, $\hat{v}_{\perp \mu}\coloneqq v_{\perp\mu}/\sqrt{|v_{\perp}|^{2}}$.
Furthermore, let us introduce the two-component spinor $c_h(q)$ as $u_{\text{R}}(q)=(c_{\text{R}}(q),\vb{0})^\mathrm{T}$ and $u_{\text{L}}(q)=(\vb{0},c_{\text{L}}(q))^\mathrm{T}$.
Then, since $c_{h}(q)$ and $c_h(p)$ are the positive- and negative-energy solutions of the Weyl equations, $(q\cdot \sigma) c_\mathrm{R}= (q\cdot \bar\sigma) c_\mathrm{L} = 0$, where $\sigma^\mu = (1,{\bm \sigma})$ and $\bar{\sigma}^\mu = (1,-{\bm \sigma})$ with $\sigma^i$ ($i=1,2,3$) being the Pauli matrices.
Eventually, from eq.~\eqref{eq:epsilon_RL}, we obtain the spinor representation of the right-handed polarization tensor:
\begin{equation}
    e^{\rmR}_{\mu\nu}(q) = \frac{1}{2}c^{\dagger}_{\rmR}(p)\sigma_{\mu}c_{\rmR}(q)c^{\dagger}_{\rmR}(p)\sigma_{\nu}c_{\rmR}(q),
\end{equation}
where the coefficient $1/2$ follows from eq.~\eqref{eq:k_mu}.

\section{Wigner functions for polarized gravitons}\label{Wigner}
\subsection{Flat spacetime}
In this section, using the quantized perturbative metric \eqref{eq:wave_expansion}, we construct the Wigner function up to $\calO(\hbar)$ in flat spacetime.
This function represents a semi-classical distribution function including the quantum interference effects.
Our approach parallels that employed for massless fermions~\cite{Hidaka:2016yjf,Hidaka:2017auj} and photons~\cite{Hattori:2020gqh}, allowing for a distinction between right/left-handed modes.

We first define the lesser propagator for right-handed gravitons as~\cite{Bellac:2011kqa}
\begin{equation}
    \label{eq:Wigner}
    W_{\mu\nu\rho\sigma}(x,q) 
  =\int \rmd^4y\, \rme^{-\rmi\frac{q\cdot y}{\hbar}}\langle h^{\rmR}_{\rho\sigma}\left(x+\tfrac{y}{2}\right)h^{\rmR}_{\mu\nu}\left(x-\tfrac{y}{2}\right)\rangle.
\end{equation}
From this definition, one can show that $W_{\mu\nu\rho\sigma}$ satisfies $W_{\mu\nu\rho\sigma} = W^{\dagger}_{\rho\sigma\mu\nu}$. 
By replacing the photon polarization vector $\epsilon_{\mu}$ with the graviton polarization tensor $e_{\mu\nu}$, the corresponding expression can be obtained from the same procedure as in ref.~\cite{Hattori:2020gqh};
details of the calculations are provided in appendix~\ref{app:Wigner}.
Finally, the Wigner function can be separated into its real and imaginary parts, 
$W_{\mu\nu\rho\sigma}=\calS_{\mu\nu\rho\sigma}+\rmi \calA_{\mu\nu\rho\sigma}$.
Each term can be calculated in terms of a frame-dependent distribution function for right-handed gravitons, $f_{\rmR}$, as
\begin{equation}\label{eq:SR}
    \begin{split}
    \calS_{\mu\nu\rho\sigma}
    &= 2\pi^2 G \delta(q^2)\text{sgn}(q\cdot n)
    \bigg\{\left(4P_{(\mu\rho}P_{\nu)\sigma}-S_{(\mu\rho}S_{\nu)\sigma}\right)\\
    &\qquad\qquad\qquad
    -\frac{\hbar}{(q\cdot n)}\qty(\hat{q}_{\perp(\mu}P_{\nu)(\rho}S_{\sigma)\alpha}+\hat{q}_{\perp(\mu}S_{\nu)(\rho}\Delta_{\sigma)\alpha}+(\mu\nu \leftrightarrow \rho\sigma))\partial^{\alpha}\bigg\}f_{\rmR}\,,
    \end{split}
\end{equation}

\begin{equation}\label{eq:AR}
    \begin{split}
    \calA_{\mu\nu\rho\sigma}
    &=-4\pi^2 G \delta(q^2)\text{sgn}(q \cdot n)\bigg\{(S_{(\mu\rho}P_{\nu)\sigma}+P_{(\mu\rho}S_{\nu)\sigma})\\
    &\qquad\qquad\qquad+\frac{\hbar}{(q \cdot n)}\qty(\hat{q}_{\perp(\mu}P_{\nu)(\rho}\Delta_{\sigma)\alpha}-\frac{1}{4}\hat{q}_{\perp(\mu}S_{\nu)(\rho}S_{\sigma)\alpha}-(\mu\nu\leftrightarrow \rho\sigma))\partial^{\alpha}\bigg\}f_{\rmR}\,.
    \end{split}
\end{equation}
The standard polarization tensor in the TT gauge and the spin tensor of gravitons are given, respectively, as
\begin{equation}
    P_{\mu\nu} = n_{\mu}n_{\nu} - \eta_{\mu\nu} - \hat{q}_{\perp\mu}\hat{q}_{\perp\nu}, \qquad S_{\mu\nu} = \frac{2\epsilon_{\mu\nu\alpha\beta}q^{\alpha}n^{\beta}}{q\cdot n}.
\end{equation}
Also, $\Delta_{\mu\nu}$ is the projection operator defined as $\Delta_{\mu\nu} \coloneqq \eta_{\mu\nu} - n_{\mu} n_{\nu}$.
Note that $\calS_{\mu\nu\rho\sigma}$ and $\calA_{\mu\nu\rho\sigma}$ are respectively symmetric and antisymmetric under the exchange $(\mu\nu)\leftrightarrow (\rho\sigma)$.

Based on eqs.~\eqref{eq:SR} and \eqref{eq:AR}, we may also write down the lesser propagator for left-handed gravitons by replacing the spin tensor $S_{\mu\nu}$ with $-S_{\mu\nu}$.
Although $W_{\mu\nu\rho\sigma}$ itself is gauge dependent, it can be utilized to calculate gauge-invariant quantities, such as the energy-momentum tensor, as discussed in the next section.
Note that, up to an overall factor, $\eta^{\mu\rho} W_{\mu\nu\rho\sigma}$ is the same as
the Wigner function for photons
since the direct product of polarization vectors $\epsilon_{\mu}$ forms the polarization tensor $e_{\mu\nu}$. 

\subsection{Curved spacetime}

Let us extend the above formulation to the graviton quantum transport theory in curved spacetime, based on the same strategy as refs.~\cite{Liu:2018xip, Mameda:2022ojk}.
The gravitational field $h_{\mu\nu}$ is now defined by, instead of eq.~\eqref{eq:h_flat},
\begin{equation}
 h_{\mu\nu} = g_{\mu\nu}-\bar{g}_{\mu\nu} ,
\end{equation}
where $\bar{g}_{\mu\nu}$ is a background metric.

While the diffeomorphism-covariant Wigner function is constructed with the covariant derivative $\nabla_{\mu}$ instead of the usual derivative $\partial_{\mu}$, its property $[\nabla_{\mu},y^\nu]\neq 0$ is disadvantageous in deriving the equation of motion for the Wigner function under the $\order{\hbar}$ expansion.
For this reason, we here introduce the horizontal lift of $\nabla_{\mu}$ onto the tangent bundle,
\begin{equation}
 D_{\mu}\Phi(x,y) = (\nabla_{\mu}-\Gamma^{\lambda}_{\mu\nu}y^{\nu}\partial_{\lambda}^{y})\Phi(x,y)
 \label{eq:derivative_y}
\end{equation}
with $\partial^{y}_{\nu} = {\partial}/{\partial y^{\nu}}$ and an arbitrary tensor field $\Phi(x,y)$.
Thanks to the property $[D_\mu, y^\nu] = 0$, the covariantly-translated gravitational field can be expressed as
\begin{equation}
 h_{\mu\nu}(x,y)
 \coloneqq \left(1 + y^\mu\nabla_\mu + \frac{1}{2}y^\mu y^\nu \nabla_\mu \nabla_\nu +\cdots \right)h_{\mu\nu}(x) 
 = \rme^{y\cdot D}h_{\mu\nu} (x) .
\end{equation}
Thus, the graviton Wigner function in curved spacetime, $W_{\mu\nu\rho\sigma}(x,q)$, can be written as
\begin{equation}
    W^{h}_{\mu\nu\rho\sigma}(x,q) = \int_{y}\rme^{-\rmi \frac{q \cdot y}{\hbar}}\langle h_{\rho\sigma}^{h}(x,\tfrac{y}{2})h^{h}_{\mu\nu}(x,-\tfrac{y}{2})\rangle,
\end{equation}
where $\int_{y} = \int \rmd^4 y \sqrt{-\bar{g}}$ and $\bar{g}=\det (\bar{g}_{\mu\nu})$.
A similar lifted derivative is also introduced for a phase-space tensor $\Phi(x,q)$ as
\begin{equation}
 D_{\mu}\Phi(x,q) = (\nabla_{\mu}+\Gamma^{\lambda}_{\mu\nu}q_{\lambda}\partial^{\nu}_{q})\Phi(x,q) ,
 \label{eq:derivative_q}
\end{equation}
where $\partial^\nu_q = \partial/\partial q_\nu$.
Note that $D_{\mu}$ satisfies the property $[D_{\mu}, q_{\nu}]=0$.

To construct the Wigner function in curved spacetime, we fix the frame ambiguity by choosing a reference vector $n_{\mu}$ subject to the parallel transport condition $\nabla_{\mu}n_{\nu}=0$.
Due to this condition, the covariant derivative $\nabla_{\mu}$ passes through the frame vector $n_{\mu}$ contained in the spin tensor $S_{\mu\nu}$ and the polarization tensor $P_{\mu\nu}$, and acts solely on the distribution function.
By replacing $\partial_{\mu}$ to $D_{\mu}$, we obtain the Wigner function in curved spacetime in the same form as in flat spacetime.

Let us write down the graviton transport equation.
We hereafter focus on the $\calO(\hbar)$ perturbation theory,
ignoring the Riemann curvature, which appears as the $\order{\hbar^2}$ contribution under our power-counting scheme.
The equation of motion for the Wigner function is 
derived in a parallel manner to that in ref.~\cite{Mameda:2022ojk} (see also appendix~\ref{app:transport_equation}).
The resulting expression is
\begin{equation}
\begin{split}\label{eq:EoM_Wigner}
\bar{g}^{\alpha\beta}\qty(q_{(\alpha}+\frac{\rmi\hbar}{2}D_{(\alpha})\qty(q_{\beta)}+\frac{\rmi\hbar}{2}D_{\beta)})W_{\mu\nu\rho\sigma}= 0\,.
    \end{split}
\end{equation}
Thanks to the hermitian property of $W_{\mu\nu\rho\sigma}$, its real and imaginary parts, $\mathcal{S}_{\mu\nu\rho\sigma}$ and $\mathcal{A}_{\mu\nu\rho\sigma}$, are written as $\mathcal{S}_{\mu\nu\rho\sigma}
    = \frac{1}{2} \left( W_{\mu\nu\rho\sigma} + W_{\rho\sigma\mu\nu} \right)$ and $\mathcal{A}_{\mu\nu\rho\sigma}
    = -\frac{\rmi}{2} \left( W_{\mu\nu\rho\sigma} - W_{\rho\sigma\mu\nu} \right)$, respectively.
Decomposing eq.~(\ref{eq:EoM_Wigner}) into the real and imaginary parts, we arrive at
\begin{equation}
    q^2 \calS_{\mu\nu\rho\sigma} = q^2 \calA_{\mu\nu\rho\sigma} = 0,
    \label{eq:on-shell}
\end{equation}
\begin{equation}
    q \cdot D \calS_{\mu\nu\rho\sigma} = q \cdot D \calA_{\mu\nu\rho\sigma} = 0\label{eq:Boltzmann_eq}.
\end{equation}
Equations~\eqref{eq:on-shell} correspond to the on-shell condition, while eqs.~\eqref{eq:Boltzmann_eq} provide the kinetic equation for the graviton distribution function $f_{h}$, 
which is up to $\calO(\hbar^0)$ equivalent to the collisionless Boltzmann equation: $\delta(q^2)q^{\mu}(\partial_{\mu}+\Gamma^{\rho}_{\mu\nu}q_{\rho}\partial^{q}_{\nu})f_{h}=0$.

We next turn to the gauge conditions for the Wigner function.
In curved spacetime, the TT gauge conditions read $\nabla^{\mu}h_{\mu\nu} = 0$, $\bar{g}^{\mu\nu}h_{\mu\nu} = 0$ and $n^{\mu}h_{\mu\nu} = 0$, which are translated into
\begin{equation}
  \qty(q^{\mu}+\frac{ \rmi \hbar}{2}D^{\mu})W_{\mu\nu\rho\sigma}=0,
  \qquad
  \bar{g}^{\mu\nu}W_{\mu\nu\rho\sigma} = 0,
  \qquad
  n^{\mu}W_{\mu\nu\rho\sigma} =0,
\end{equation}
or equivalently,
\begin{gather}
    q^{\mu}\calS_{\mu\nu\rho\sigma} - \frac{\hbar}{2}D^{\mu}\calA_{\mu\nu\rho\sigma} = q^{\mu}\calA_{\mu\nu\rho\sigma} + \frac{\hbar}{2}D^{\mu}\calS_{\mu\nu\rho\sigma} = 0,
    \label{eq:TT_SA1}\\
    \bar{g}^{\mu\nu}\mathcal{S}_{\mu\nu\rho\sigma} =
\bar{g}^{\mu\nu}\mathcal{A}_{\mu\nu\rho\sigma} = 0,
\label{eq:TT_SA2}\\
    n^{\mu}\mathcal{S}_{\mu\nu\rho\sigma} = 
n^{\mu}\mathcal{A}_{\mu\nu\rho\sigma} = 0.
\label{eq:TT_SA3}
\end{gather}
These conditions eliminate the unphysical degrees of freedom in the Wigner function.

Similarly to the cases of chiral fermions and photons, the graviton Wigner function in the TT gauge may be identified as the same form as the flat-spacetime solutions in eqs.~\eqref{eq:SR} and~\eqref{eq:AR}, but with the replacement $\partial_\mu \to D_\mu$ and $\eta_{\mu\nu}\to \bar{g}_{\mu\nu}$.
Indeed, one can show that such Wigner functions, $\mathcal{S}_{\mu\nu\rho\sigma}$ and $\mathcal{A}_{\mu\nu\rho\sigma}$, fulfill the constraint conditions in eqs.~\eqref{eq:on-shell} and eqs.~\eqref{eq:TT_SA1}--\eqref{eq:TT_SA3}.
The same argument is also utilized to obtain the Wigner function of photons in general coordinates~\cite{Mameda:2022ojk}:
in the rigid rotating coordinate, this photonic Wigner function correctly reproduces the zilch vortical effect~\cite{Chernodub:2018era,Copetti:2018mxw} and photonic chiral vortical effect~\cite{Avkhadiev:2017fxj,Yamamoto:2017uul,PhysRevA.96.043830,Huang:2018aly,Prokhorov:2020okl}.

\section{Graviton energy-momentum tensor}\label{graviton_current}
In this section, we perform the Wigner transformation of the energy-momentum tensor for gravitons and derive anomalous currents in a vorticity and in a weak-gravity metric $\bar{g}_{\mu\nu}$.
In practice, thermal equilibrium for gravitons is unlikely to be realized in nature, since the rates of processes that would establish equilibrium are smaller than the cosmic expansion rate set by the Hubble parameter~\cite{Weinberg:1972kfs}. 
Nevertheless, for illustrative purposes, we assume thermal equilibrium, i.e., the Bose-Einstein distribution, and verify that the energy-momentum tensor $T_{\mu\nu}^{h}$ is independent of the choice of the frame vector. We also derive the spin Hall effect of gravitons and demonstrate that its amplitude is twice that of the corresponding effect for photons.

For $g_{\mu\nu} = \bar{g}_{\mu\nu} + h_{\mu\nu}$, the full Einstein tensor $\mathcal{G}_{\mu\nu} =R_{\mu\nu} -\frac{1}{2}{g}_{\mu\nu}R$ is also expanded in terms of $h_{\mu\nu}$ as $\mathcal{G}_{\mu\nu}=\mathcal{G}_{\mu\nu}^{[0]}+\mathcal{G}_{\mu\nu}^{[1]}+\mathcal{G}_{\mu\nu}^{[2]}+\cdots$, where $[\cdot]$ denotes the power of $h_{\mu\nu}$ involved there.
Then, the Einstein equation in the absence of matter field leads to $\mathcal{G}_{\mu\nu}^{[0]} = - (\mathcal{G}_{\mu\nu}^{[1]} +\mathcal{G}_{\mu\nu}^{[2]}+\cdots )$.
Compared with the standard form of the Einstein equation, the right-hand side can be regarded as the energy-momentum tensor of the gravitational field $h_{\mu\nu}$, implying
\begin{equation}
  \mathcal{T}_{\mu\nu} = - \frac{1}{8\pi G}\left(\mathcal{G}_{\mu\nu}^{[1]}+\mathcal{G}_{\mu\nu}^{[2]} \right),
\end{equation}
up to $\order{|h_{\mu\nu}|^2}$.
Furthermore, under the TT gauge condition, the above equation is reduced~\cite{Altas:2019qcv}: one can show that $\calG^{[1]}_{\mu\nu} = 0$ and $\calG^{[2]}_{\mu\nu} = R_{\mu\nu}^{[2]} -\frac{1}{2}\bar{g}_{\mu\nu}R^{[2]}$ with
\begin{equation}
    \begin{split}
        R_{\mu\nu}^{[2]}
        &= -\frac{1}{2}h^{\sigma\beta}\nabla_{\sigma}\left(\nabla_{\nu}h_{\mu\beta}+\nabla_{\mu}h_{\nu\beta}-\nabla_{\beta}h_{\mu\nu}\right)
        +\frac{1}{4}\nabla_{\nu}\nabla_{\mu}(h_{\alpha\beta}h^{\alpha\beta})\\
        &\quad
        -\frac{1}{4}\nabla_{\nu}h^{\alpha\beta}\nabla_{\mu}h_{\alpha\beta}
        -\frac{1}{2}\nabla^{\sigma}h_{\mu\alpha}\nabla^{\alpha}h_{\nu\sigma}
        +\frac{1}{2}\nabla^{\sigma}h_{\mu\alpha}\nabla_{\sigma}h_{\nu}^{\alpha}
    \end{split}
\end{equation}
and
\begin{equation}\label{secondorderRicciscalor}
    R^{[2]}=\nabla_{\sigma}\nabla_{\lambda} \qty(\frac{3}{8}\bar{g}^{\sigma\lambda}h^{2}_{\alpha\beta}-\frac{1}{2}h^{\lambda}_{\rho}h^{\rho\sigma}).
\end{equation}

In order to identify the Wigner-function form of the above $\mathcal{T}_{\mu\nu}$, we symmetrize the indices of the two $h_{\mu\nu}$ fields involved in each term, convert $\nabla_\mu$ to $D_\mu$, and replace them with $h_{\mu\nu}(x,\pm \frac{y}{2})$;
e.g.,
\begin{equation}
    h_{\rho\sigma}\nabla_{\mu}\nabla_{\nu}h_{\lambda\eta} \rightarrow \frac{1}{2}\Bigl[\left\{D_{\mu}D_{\nu}h_{\lambda\eta}\left(x, \tfrac{y}{2}\right)\right\}h_{\rho\sigma}\left(x,- \tfrac{y}{2}\right)+h_{\rho\sigma}\left(x, \tfrac{y}{2}\right)\left\{D_{\mu}D_{\nu}h_{\lambda\eta}\left(x,- \tfrac{y}{2}\right)\right\}\Bigr].
\end{equation}
After performing the Wigner transformation, the phase-space representation of $\mathcal{T}_{\mu\nu}(x,q)$ becomes
\begin{equation}
\begin{split}
        &\mathcal{T}_{\mu\nu}(x,q)\\
        &= -\frac{1}{16\pi \hbar^2 G} \left[ 
        \frac{1}{2}
            \qty(
                \calD_{\sigma}\calD_{\nu}W_{\mu\beta}^{\;\;\;\;\sigma\beta}
                +\calD_{\sigma}^{\ast}\calD_{\nu}^{\ast} W^{\sigma\beta}_{\;\;\;\;\mu\beta}
                )
        +\frac{1}{2}
            \qty(
                \calD_{\sigma}\calD_{\mu}W_{\nu\beta}^{\;\;\;\;\sigma\beta}
                +\calD_{\sigma}^{\ast} \calD_{\mu}^{\ast} W^{\sigma\beta}_{\;\;\;\;\nu\beta}
                )\right.\\
        &\quad
        -\frac{1}{2}
            \qty(
                \calD_{\sigma}\calD_{\beta} W^{\;\;\;\;\sigma\beta}_{\mu\nu}
                +\calD_{\sigma}^{\ast}\calD_{\beta}^{\ast} W^{\sigma\beta}_{\;\;\;\;\mu\nu}
                )
        -\frac{1}{2}
            \qty( 
                \calD_{\nu}\calD_{\mu}
                +\calD_{\nu}^{\ast}\calD_{\mu}^{\ast}
                )  W_{\alpha\beta}^{\;\;\;\;\alpha\beta}\\
        &\quad
        +\frac{1}{4}
            \qty(
                \calD_{\nu}^{\ast}\calD_{\mu}
                +\calD_{\mu}^{\ast}\calD_{\nu}) W_{\alpha\beta}^{\;\;\;\;\alpha\beta}
        -\frac{1}{2}
            \qty(
                \calD_{\sigma}^{\ast}\calD_{\alpha} W^{\alpha\;\;\sigma}_{\;\;\mu\;\;\nu}
                +\calD_{\alpha}^{\ast}\calD_{\sigma} W_{\mu\;\;\nu}^{\;\;\alpha\;\;\sigma}
                )\\
        &\quad
        +\frac{1}{2}\calD_{\sigma}^{\ast}\calD^{\sigma}
                (W_{\;\;\nu\mu\alpha}^{\alpha}
                +W_{\mu\alpha\;\;\nu}^{\;\;\;\;\alpha})
        -\frac{3}{8}\bar{g}_{\mu\nu}
                 (\calD_{\lambda}^{\ast}\calD^{\lambda}
                 +\calD_{\lambda}\calD^{\ast\lambda}) W_{\alpha\beta}^{\;\;\;\;\alpha\beta}\\
        &\quad\left.
        -\frac{1}{4} \bar{g}_{\mu\nu}
            \qty(
            \calD^{\ast}_{\sigma}\calD_{\lambda} W^{\rho\sigma\lambda}_{\;\;\;\;\;\;\rho}
            +\calD^{\ast}_{\lambda}\calD_{\sigma} W^{\lambda\;\;\rho\sigma}_{\;\;\rho})\right],
\end{split}
\end{equation}
where $\calD_{\mu} = q_{\mu} + \frac{\rmi \hbar}{2}D_{\mu}$. 
The graviton energy-momentum tensor $T_{\mu\nu}$ is the momentum integral of $\mathcal{T}_{\mu\nu}$. 
Hereafter, we explicitly write the subscript $h=\rmR,\rmL$, corresponding to the signs $+$ and $-$, respectively.
Plugging the solutions of $W^{h}_{\mu\nu\rho\sigma}$, we hence find 
\begin{equation}\label{eq:T}
    \begin{split}
        T^{h}_{\mu\nu}
        =\int_{q}\mathcal{T}^{h}_{\mu\nu}
        = \frac{\pi}{\hbar^2}\int_{q}\delta(q^2)\text{sgn}(q_{0})(2q_{\mu}q_{\nu}\pm\hbar q_{(\mu}S_{\nu)\alpha}D^{\alpha})f^{h}(x,q),
    \end{split}
\end{equation}
where $\int_{q}=\int \frac{\rmd ^4 q}{(2\pi)^4}\frac{1}{\sqrt{-\bar{g}}}$.

Below we will show that $T_{\mu\nu}$ is independent of the choice of frame vector $n_{\mu}$. 
As in the case of fermions and photons~\cite{Liu:2018xip, Mameda:2022ojk}, the equilibrium distribution incorporating spin-vorticity coupling should be
\begin{equation}\label{equilibrium}
    f_{h} = N(g_{h}), \quad g_{h}=q\cdot U \pm \frac{\hbar}{2}S^{\mu\nu}\nabla_{\mu}U_{\nu},
\end{equation}
where $N(x)=(\rme^x -1)^{-1}$ and $U^{\mu} = \beta u^{\mu}$, with $\beta=1/T$ the inverse temperature and $u^{\mu}$ the fluid four velocity satisfying the Killing condition
\begin{equation}\label{Killing}
    \nabla_{(\lambda}U_{\alpha)}=0.
\end{equation}
The $\calO(\hbar)$ component $T^{h(1)}_{\mu\nu}$ appearing in eq.~\eqref{eq:T} can be written as
\begin{equation}\label{eq:T1}
    \begin{split}
        T^{h(1)}_{\mu\nu}&= \pm\frac{\pi}{\hbar^2}\int_{q}\delta(q^2)\text{sgn}(q_{0})\cdot \hbar(q_{\beta}q_{(\mu}S_{\nu)\alpha}\nabla^{\alpha}U^{\beta}+q_{\mu}q_{\nu}S^{\alpha\beta}\nabla_{\alpha}U_{\beta})N'(q\cdot U)\\
        &=\pm\frac{\pi}{\hbar^2}\int_{q}\delta(q^2)\text{sgn}(q_{0})\cdot \hbar \nabla^{\alpha}U^{\lambda}\epsilon_{\alpha\lambda\beta(\mu}q_{\nu)}q^{\beta}N'(q\cdot U)\,,
    \end{split}
\end{equation}
where the superscript $(\cdot)$ denotes the order of $\hbar$.
Here, we used eq.~\eqref{Killing} and the relation
\begin{equation}
    S_{\nu[\alpha}q_{\lambda]}=-S_{\alpha\lambda}q_{\nu}-2\epsilon_{\alpha\lambda\nu\beta}\qty(q^{\beta}-\frac{q^2n^{\beta}}{q\cdot n})\,,
\end{equation}
which follows from the Schouten identity. Of course, the $\calO(\hbar^0)$ component of the energy-momentum tensor is also independent of the frame vector $n^{\mu}$.
Therefore, we confirm that the total energy-momentum tensor is independent of the frame vector $n^{\mu}$ up to $\calO(\hbar)$.

As a theoretically simple case, let us calculate the energy current in a slowly rotating coordinate system. The metric compatible with the frame vector $n_{\mu} =(1,\vb{0})$ is taken to be
\begin{equation}\label{eq:g_rot}
    \bar{g}_{\mu\nu} = \eta_{\mu\nu} + \gamma_{\mu\nu},\qquad |\gamma_{\mu\nu}|\ll 1,\qquad \gamma_{0i} =\gamma _{0i}(\bx), \qquad \gamma_{00}=\gamma_{ij} =0.
\end{equation}
Note that the background metric $\bar{g}_{\mu\nu}$ is expanded in terms of the perturbation $\gamma_{\mu\nu}$, which is independent of the metric perturbation $h_{\mu\nu}$.
In the following, we work up to $\calO(|\gamma_{\mu\nu}|)$.
We adopt the equilibrium state~\eqref{equilibrium} with $u^{\mu} = \delta^{\mu}_{0}$, such that $u_{i} = \gamma_{i0}(\bx)$ and $\beta$ is a constant. 
From eq.~\eqref{eq:T1}, the positive-energy contribution to the chiral vortical current is
\begin{equation}\label{eq:CVE}
    \begin{split}
        J^{h\;\text{vort}}_{i} &= \pm \frac{\pi}{\hbar}\int_{q} \frac{\delta(q_{0} - |\bq|)}{2|\bq|}  \nabla^{\alpha}U^{\lambda}\epsilon_{\alpha\lambda\beta(0}q_{i)}q^{\beta}N'(q \cdot U)\\
        &= \mp \frac{8\beta\omega_{i}}{3\hbar}\int_{\bq} |\bq| N'(\beta|\bq|)
        = \pm \frac{2\zeta(3)}{\pi^2\hbar}T^{3} \omega_{i},
    \end{split}
\end{equation}
where $\omega^{\mu} = \frac{1}{2} \epsilon^{\mu\nu\rho\sigma}u_{\nu}\nabla_{\rho}u_{\sigma}$ is the covariant fluid vorticity and $\int_{\bq} \coloneq \int \frac{\rmd^3 \bq}{(2\pi)^3}\frac{1}{\sqrt{-\bar{g}}}$.
As we mentioned in the introduction, the correct physical dimension can be recovered by multiplying $\hbar^2$.

On the other hand, the weak background gravity metric, 
\begin{equation}\label{SHE_metric}
    \bar{g}_{\mu\nu}=\eta_{\mu\nu}+\gamma_{\mu\nu}, \quad|\gamma_{\mu\nu}|\ll1, \quad \gamma_{ij}=4\delta_{ij}\phi(\bx), \quad \gamma_{0\mu}=0,
\end{equation}
admits $u^{\mu} = (1, \bu)$ as a Killing vector, if the following conditions are fulfilled:
\begin{equation}
    \begin{split}
        \partial_{(i}u_{j)} = 0 , \qquad 
        u^{i}\partial_{i} \phi = 0.
    \end{split}
\end{equation}
In this case, the energy current can be written in the same form as in eq.~\eqref{eq:CVE}, except that the fluid-vorticity term is shifted by the Lorentz transformation according to
\begin{equation}
    \begin{split}
        \bomega = \tilde{\bomega} - 2\bnabla \phi \times \bu,
    \end{split}
\end{equation}
where $\tilde{\bomega} \coloneqq \frac{1}{2}(1 + 2\phi) \bnabla \times \bu$.
The first term leads to the current $\tilde{\bJ}^{h\;\text{vort}}$, corresponding to eq.~\eqref{eq:CVE}.
The second term leads to the spin Hall energy current, which is obtained by replacing $\omega_{i}$ in eq.~\eqref{eq:CVE} with $-2 (\bnabla \phi \times \bu)_{i}$: 
\begin{equation}\label{eq:SHE}
    \begin{split}
        J_{i}^{h\;\text{Hall}} = \mp \frac{4\zeta(3)}{\pi^2\hbar}T^{3} (\bnabla \phi \times \bu)_{i}.
    \end{split}
\end{equation}
The energy currents for $h =\rmR, \rmL$ point in the opposite directions. Compared with the corresponding photonic current, the magnitude is larger by a factor of two.

\section{Kinetic equation}\label{kinetic_eq}
In this section, we identify the equation of motion for a graviton from the graviton energy-momentum tensor obtained in the previous section.
In kinetic theory, the energy current is obtained by multiplying the particle velocity by its energy and distribution function, and then integrating over momentum space.
Here, we reverse this logic to infer the equation of motion in a general metric, including the metric~\eqref{SHE_metric}. In this way, by identifying the presence of the Berry curvature in the quantum correction term, we confirm consistency with the previous work~\cite{Yamamoto:2017gla}.

To this end, we first solve the on-shell condition $\bar{g}^{\mu\nu}q_{\mu}q_{\nu}=0$, leading to the energy-dispersion relation
\begin{equation}
    q_0 = E_{\bm{q}} \coloneq 
    \frac{1}{\bar{g}^{00}}\left(-\bar{g}^{0i}q_{i}
    +\sqrt{(\bar{g}^{0i}q_{i})^2-\bar{g}^{00} 
    \bar{g}^{ij} 
    q_{i}q_{j}}\right).
\end{equation}
Also we note that under this on-shell condition, $q_0$ depends on $\bq$ and $x$ as $\partial_\mu q_0 = \Gamma^\rho_{\mu\nu}q^\nu q_\rho/q_0$ and $\partial_q^i q_0 = -q^i/q^0$.
Thus, after the integration over $q_0$ taking into account the on-shell condition, the derivative operators
\begin{equation}
\tilde{\partial}_{\mu} \coloneqq \partial_{\mu} + \frac{1}{q_{0}}\Gamma_{\mu\nu}^{\rho}q^{\nu}q_{\rho} \partial_q^0, 
\qquad
\tilde{\partial}_{q}^{i} \coloneqq \partial^{i}_{q} - \frac{q^{i}}{q^{0}}\partial_{q}^{0}
\end{equation}
are replaced with $\partial_\mu$ and $\partial_q^i$, respectively.
Then, by focusing on the positive-energy states and integrating over $q_{0}$, the energy current in general metric from eq.~\eqref{eq:T} is expressed as
\begin{equation}
    \begin{split}\label{eq:energy_current}
        J^{h}_{i}\coloneqq T^{h}_{0i} 
        &= \frac{\pi}{2\hbar^2}\int _{q}\frac{\delta(q_{0}-E_{\bq})}{\tilde{E}_{\bq}}
        \qty[2q_{0}q_{i} \pm \hbar q_{(0}S_{i)}^{\;\;\alpha}(\tilde{\partial}_{\alpha}+\Gamma_{\alpha i}^{\lambda}q_{\lambda}\tilde{\partial}_{q}^{i})]f^{h}(x,q)\\
        &=\frac{1}{4\hbar^2}\int_{\bq}\frac{E_{\bq}}{\tilde{E}_{\bq}}\qty[2q_{i}\pm 2\hbar \frac{\epsilon_{i}^{\;\;kl0}q_{l}
        }{\tilde{E}_{\bq}}\qty(\partial_{k}+ \Gamma_{k j}^{0}E_{\bq}\partial_{q}^{j}+\Gamma_{k j}^{m}q_{m}\partial_{q}^{j})]f^{h}(x,E_{\bq},\bq)\,,
    \end{split}
\end{equation}
where $\tilde{E}_{\bq}=|\bar{g}^{00}E_{\bq} + \bar{g}^{0i}q_{i}|$.
The energy current obtained from the Wigner function includes, in the second term, a magnetization current involving the spatial derivative, which originates from the spin of the particle~\cite{Son:2012zy,Chen:2014cla}.
To identify the single-particle equation of motion, we separate this magnetization current from the total current and associate only the transport part with $\dot{\bx}$.
The remaining parts in $\calO(\hbar)$ then determine
\begin{equation}
    \begin{split}
        J^{h (1)}_{i\;\text{trans}} &=\pm \frac{1}{4\hbar^2}\int_{\bq}2\hbar\frac{E_{\bq}}{\tilde{E}_{\bq}}\qty[ \frac{\epsilon_{i}^{\;\;kl0}q_{l}
        }{\tilde{E}_{\bq}}\qty( \Gamma_{k j}^{0}E_{\bq}\partial_{q}^{j}+\Gamma_{k j}^{m}q_{m}\partial_{q}^{j})]f^{h}(x,E_{\bq},\bq)\\
        &= \mp\frac{1}{4\hbar^2} \int_{\bq} 2\hbar\partial^{j}_{q}\qty[\frac{E_{\bq}\epsilon_{i}^{\;\;kl0}q_{l}}{\tilde{E}^{2}_{\bq}} \qty(\Gamma_{k j}^{0}E_{\bq} + \Gamma_{k j}^{m}q_{m})]f^{h}(x,E_{\bq}, \bq)\,,
    \end{split}
\end{equation}
where we integrated by parts with respect to momentum in the second line. The equation of motion for a graviton in the general background gravity reads
\begin{equation}\label{eq:EOM}
    \begin{split}
        \dot{x}_{i}^{h} &\simeq \frac{1}{\tilde{E}_{\bq}}\qty(q_{i} \mp \hbar \partial^{j}_{q}\qty[\frac{E_{\bq}\epsilon_{\;\;\;\;\;i}^{0kl}q_{l}}{\tilde{E}^{2}_{\bq}} \qty(\Gamma_{k j}^{0}E_{\bq} + \Gamma_{k j}^{m}q_{m})]).
    \end{split}
\end{equation}
Note that the velocity $\dot{\bx}$ depends on the frame vector $n^{\mu}$.
However, the energy current is independent of the choice of $n^{\mu}$, because the frame dependence of $\dot{\bx}$ is exactly compensated by that of the distribution function (and the magnetization current).

For the metric~\eqref{SHE_metric}, 
the equation of motion (\ref{eq:EOM}) reduces to
\begin{equation}\label{eq:EOM_SHE}
    \begin{split}
        \dot{x}^{h}_{i} &\simeq \frac{1}{|\bq|(1 + 2\phi)}\qty[q_{i} \pm \hbar \partial^{j}_{q}\qty(\frac{2}{|\bq|}(1 -2\phi)\left[q_{j}(\bnabla\phi \times \hat{\bq})_{i} - \epsilon^{0}_{\;\;jli}q^{l}(\bq \cdot \bnabla\phi)\right])]\\
        &=(1 -2\phi)\hat{q}_{i} \mp 4\hbar\left(\bnabla\phi \times \frac{\hat{\bq}}{|\bq|}\right)_{i}\,.
    \end{split}
\end{equation}
The ${\cal O}(\hbar)$ correction is consistent with the equation of motion for a graviton derived in ref.~\cite{Yamamoto:2017gla},%
\footnote{On the other hand, the sign of $2\phi$ at ${\cal O}(\hbar^0)$ is different. This is because, although the refractive index,  $n=\sqrt{g_{ii}/g_{00}}$ (with no summation over $i$), is the same for the metric (\ref{SHE_metric}) and for that in ref.~\cite{Yamamoto:2017gla}, the $g_{ii}$ component itself differs between the two.}
\begin{equation}
\dot {\bm x}^{(1)} = \dot {\bm q}\times \hbar{\bm \Omega}_{\bm q}, \qquad
 \dot {\bm q}  = -2|{\bm q}| {\bm \nabla} \phi,
\end{equation}
where ${\bm \Omega}_{\bm q}=\pm{2\hat{\bq}}/{|\bq|^2}$ is the Berry curvature for right- and left-handed gravitons. This quantum correction to the graviton equation of motion in weak gravitational fields is twice as large as that for photons~\cite{Mameda:2022ojk}, reflecting the graviton helicities $\pm2$.

\section{Conclusion and discussion}\label{conclusion_discussion}
In this paper, we developed a quantum kinetic theory for gravitons in curved spacetime from quantum field theory, following the methodology presented for photons in refs.~\cite{Hattori:2020gqh, Mameda:2022ojk}. 
Using the graviton Wigner function, we derived the chiral vortical effect in a rotating coordinate system and spin Hall effect in a weak background gravitational field, both induced by the Berry curvature of gravitons. 
We also obtained generally covariant currents, which is a clear advantage over the noncovariant formulation of ref.~\cite{Yamamoto:2017gla}.
The resulting energy current is twice as large as the photonic one in ref.~\cite{Mameda:2022ojk}.
It should be emphasized that, although we adopted a specific frame vector to define the spin polarization of gravitons, our results do not depend on the choice of the frame vector, as in the fermionic~\cite{Chen:2015gta, Hidaka:2017auj} and photonic~\cite{Hattori:2020gqh} cases.

There remain several directions for future works.
It is known that the Berry curvature and the spin Hall effect of light in single-particle dynamics can be derived even within purely classical Maxwell theory~\cite{Onoda:2006ne, Nishida:2026bzu}. This suggests that the Berry curvature and spin Hall effect of  gravitational waves, as described by the equation of motion (\ref{eq:EOM_SHE}), might also admit a derivation within purely classical general relativity. Establishing whether this is indeed the case would be an interesting theoretical question.
Closely related to this issue, previous studies~\cite{Kubota:2023dlz, Frolov:2025bva, Kubota:2024zkv}, based on classical methods, pointed out that the trajectories of gravitational waves around a Kerr black hole depend on their helicity. An important open question is whether our analysis and those studies can be understood in a unified and mutually consistent manner.
On the other hand, graviton currents in thermal equilibrium, such as those associated with the chiral vortical effect in eq.~(\ref{eq:CVE}) and the spin Hall effect in eq.~(\ref{eq:SHE}), are manifestations of the quantum nature of gravitons, even though it is presumably difficult to expect a realistic situation in which gravitons thermalize.

For massless chiral fermions with helicity $\pm1/2$~\cite{Mameda:2025rfn} and photons with helicity $\pm 1$~\cite{Akiba:2026den}, it has recently been shown that, at ${\cal O}(\hbar^2)$, another quantum-geometric quantity, called the quantum metric~\cite{Provost1980}, appears in addition to the Berry curvature. In the present work, we have restricted ourselves to the expansion up to ${\cal O}(\hbar)$. Extending the analysis to ${\cal O}(\hbar^2)$ should make it possible to derive the corresponding quantum metric for gravitons as well. 
One may also be able to derive the energy-momentum tensor induced by the gravitational Riemann curvature at ${\cal O}(\hbar^2)$, similarly to the case of chiral fermions~\cite{Hayata:2020sqz}.

\begin{acknowledgments}
 We thank Asuka~Ito, Bei-Lok~Hu, Masahide~Yamaguchi, and Di-Lun~Yang for valuable discussions.
 This work was supported by JGC-S Scholarship Foundation (R.\,I.), and JSPS KAKENHI Grant Numbers 24K17052 (K.\,M.) and JP24K00631 (N.\,Y.).
\end{acknowledgments}

\appendix

\section{Derivation of Wigner functions for polarized gravitons}\label{app:Wigner}
In this appendix, we present a detailed derivation of the Wigner function for right- and left-handed gravitons.
This construction is essentially similar to that in ref.~\cite{Hattori:2020gqh}, except for the difference of the tensor structure between photons and gravitons. We focus on the right-handed Wigner function:
\begin{equation}
  \begin{split}
  W^{\rmR}_{\mu\nu\rho\sigma}(x,q) &= \int \rmd^4 y\, \rme^{-\rmi\frac{ q\cdot y}{\hbar}} \langle h^{\rmR}_{\rho\sigma}(x+\tfrac{y}{2})h^{\rmR}_{\mu\nu}(x-\tfrac{y}{2}) \rangle\\
  &= 32\pi G\int \rmd^4 y\, \rme^{-\rmi\frac{ q\cdot y}{\hbar}} \int \frac{\rmd^3 p'}{(2\pi)^3} \frac{1}{\sqrt{2|\bp'|}}\int \frac{\rmd^3 p}{(2\pi)^3}\frac{1}{\sqrt{2|\bp|}}\times\\ 
    & \quad \qty(N_{\bp',\bp}^{\rmR} e^{\rmR\dagger}_{\mu\nu}(p) e^{\rmR}_{\rho\sigma}(p')\rme^{\rmi \frac{p_{-}\cdot x - p_{+}\cdot y}{\hbar}} + N^{\rmR\dagger}_{\bp,\bp'} e^{\rmR}_{\mu\nu}(p)e^{\rmR\dagger}_{\rho\sigma}(p')\rme^{-\rmi\frac{p_{-}\cdot x - p_{+} \cdot y}{\hbar}}),\,
  \end{split}
\end{equation}
where $N_{\bp',\bp}^{\rmR} = \langle a^{\rmR\dagger}_{\bp'}a_{\bp}^{\rmR}\rangle$, $p_{+}=(p+p')/2$ and $p_{-}=p-p'$.

After performing the integration over $y$ and changing variables from $p$ and $p'$ to $p_{+}$ and $p_{-}$, we obtain
\begin{equation}\label{eq:W1}
    \begin{split}
        W^{\rmR}_{\mu\nu\rho\sigma}(x,q) 
        &= 16\pi G \cdot 2\pi\int \frac{\rmd^3p_{-}}{(2\pi)^3}\,\rme^{-\rmi\frac{ p_{-}\cdot x}{\hbar}}\int \rmd^3 p_{+}
            \frac{1}{\sqrt{|\bp_{+}-\frac{\bp_{-}}{2}||\bp_{+}+\frac{\bp_{-}}{2}|}}\\
        &\quad\times \left[N^{\rmR}_{\bp_{+}-\frac{\bp_{-}}{2},\bp_{+}+\frac{\bp_{-}}{2}} e_{\mu\nu}^{\rmR}\qty(p_{+}+\frac{p_{-}}{2})e^{\rmR\dagger}_{\rho\sigma}\qty(p_{+}-\frac{p_{-}}{2})\delta(p_{+}-q)\right.\\
            &\qquad\left.+N^{\rmR\dagger}_{\bp_{+}-\frac{\bp_{-}}{2},\bp_{+}+\frac{\bp_{-}}{2}} e^{\rmR\dagger}_{\mu\nu}\qty(p_{+}-\frac{p_{-}}{2})e^{\rmR}_{\rho\sigma}\qty(p_{+}+\frac{p_{-}}{2})\delta(p_{+}+q)\right],
    \end{split}
\end{equation}
where
\begin{equation}
    p^{0}_{+} =\frac{1}{2}\qty(\qty|\bq + \frac{\bp_{-}}{2}|+\qty|\bq-\frac{\bp_{-}}{2}|), \qquad p^{0}_{-}=\qty|\bq+\frac{\bp_{-}}{2}|-\qty|\bq-\frac{\bp_{-}}{2}|.
\end{equation}
To evaluate the $p_{-}$ integral analytically, we expand the integrand in powers of $p_{-}$ and keep terms up to $\calO(p_-)$, e.g., 
\begin{equation}
    \begin{split}
        e^{\text{R}}_{\mu\nu}\qty(q+\frac{p_{-}}{2})e^{\text{R}\dagger}_{\rho\sigma}\qty(q-\frac{p_{-}}{2}) =\Pi_{\mu\nu\rho\sigma}(q)+\frac{p^{\alpha}_{-}}{2}\Sigma_{\mu\nu\rho\sigma\alpha}(q)+\calO(p^2_{-}),
    \end{split}
\end{equation}
where
\begin{equation}\label{eq:Pi0Pi1}
    \begin{split}
        \Pi_{\mu\nu\rho\sigma}(q)&\coloneqq e^{\text{R}}_{\mu\nu}(q)e^{\text{R}\dagger}_{\rho\sigma}(q)=\epsilon^{\text{R}}_{\mu}\epsilon^{\text{R}}_{\nu}\epsilon^{\text{R}\dagger}_{\rho}\epsilon^{\text{R}\dagger}_{\sigma}=\frac{1}{2}\tilde{\Pi}_{(\mu\rho}\tilde{\Pi}_{\nu)\sigma},\\
        \quad\Sigma_{\mu\nu\rho\sigma\alpha}(q)&\coloneqq (\partial_{\alpha}^{q}e^{\text{R}}_{\mu\nu}(q))e^{\text{R}\dagger}_{\rho\sigma}(q)-e^{\text{R}}_{\mu\nu}(q)(\partial_{\alpha}^{q}e^{\text{R}\dagger}_{\rho\sigma}(q))\\
        &=\frac{1}{2}\qty(\tilde{\Sigma}_{(\mu\rho\alpha}\tilde{\Pi}_{\nu)\sigma}+\tilde{\Sigma}_{(\nu\sigma\alpha}\tilde{\Pi}_{\mu)\rho}),
    \end{split}
\end{equation}
with $\tilde{\Pi}_{\mu\nu}=\epsilon^{\rmR}_{\mu}\epsilon^{\rmR\dagger}_{\nu}$ and $\tilde{\Sigma}_{\mu\nu\alpha}=(\partial_{\alpha}^{q}\epsilon^{\rmR}_{\mu})\epsilon^{\rmR\dagger}_{\nu}-\epsilon^{\rmR}_{\mu}(\partial_{\alpha}^{q}\epsilon^{\rmR\dagger}_{\nu})$.

Using the expansions $p^{0}_{+} = |\bq| +\calO(|\bp_{-}|^2)$ and $p^{0}_{-}= \bq\cdot \bp_{-}/|\bq| +\calO(|\bp_{-}|^2)$, we find
\begin{equation}\label{eq:W2}
    \begin{split}
        &W^{\text{R}}_{\mu\nu\rho\sigma}(x,q)\\
        &\simeq 32\pi^2 G \int \frac{\rmd^3 p_{-}}{(2\pi)^3|\bq|}\rme^{-\rmi\frac{ p_{-}\cdot x}{\hbar}}\left[\delta(|\bq|-q_{0})\qty(\Pi_{\mu\nu\rho\sigma}(q)+\frac{p^{\alpha}_{-}}{2}\Sigma_{\mu\nu\rho\sigma\alpha}(q))N^{\rmR}_{\bq-\frac{\bp_{-}}{2},\bq+\frac{\bp_{-}}{2}}\right.\\
        &\qquad\qquad \qquad\qquad \left.+\delta(|\bq|+q_{0})\qty(\Pi_{\rho\sigma\mu\nu}(-q)+\frac{p^{\alpha}_{-}}{2}\Sigma_{\rho\sigma\mu\nu\alpha}(-q))N^{\rmR\dagger}_{-\bq-\frac{\bp_{-}}{2},-\bq+\frac{\bp_{-}}{2}}\right]\\
        &=64\pi^2 G \delta(q^2)\left[\theta(q_0)\qty(\Pi_{\mu\nu\rho\sigma}(q)+\frac{\rmi\hbar}{2}\Sigma_{\mu\nu\rho\sigma\alpha}(q)\partial^{\alpha})\right.\\
        &\left.\qquad\qquad\qquad\qquad\qquad-\theta(-q_{0})\qty(\Pi_{\rho\sigma\mu\nu}(-q)+\frac{\rmi\hbar}{2}\Sigma_{\rho\sigma\mu\nu\alpha}(-q)\partial^{\alpha})\right]\check{f}_{\text{R}}(x,q)\,.
    \end{split}
\end{equation}
Here, $\check{f}_{\rmR}(x,q)$ is the distribution function for right-handed gravitons, defined for the four-momentum $q^{\mu}$, as
\begin{equation}
    \check{f}_{\text{R}}(x,q) = \left\{ \begin{aligned}
        &\check{f}_{\text{R}}(x,\bq)  \qquad\qquad\qquad(q_{0}=|\bq|)\\
        &-[1+\check{f}_{\text{R}}(x,-\bq)] \qquad(q_{0}=-|\bq|),
    \end{aligned}\right.
\end{equation}
where
\begin{equation}
    \check{f}_{\text{R}}(x,\bq) \coloneqq \int \frac{\rmd^3 p_{-}}{(2\pi)^3}N^{\rmR}_{\bq-\frac{\bp_{-}}{2},\bq+\frac{\bp_{-}}{2}}\rme^{-\rmi\frac{ p_{-}\cdot x}{\hbar}}\,.
\end{equation}
The term $\theta(-q_0)$ characterizes the outgoing gravitons, where  $\theta(x)$ denotes the step function. 

The explicit forms of $\tilde{\Pi}_{\mu\nu}$ and $\tilde{\Sigma}_{\mu\nu\alpha}$, based on the spinor-helicity representation of $e^{h}_{\mu\nu}$, are given in ref.~\cite{Hattori:2020gqh} as
\begin{align}
        \tilde{\Pi}_{\mu\nu}(q)
        &= \frac{1}{2}\qty(P_{\mu\nu}-\frac{\rmi}{2}S_{\mu\nu})\,,
        \\
        \tilde{\Sigma}_{\mu\nu\alpha}(q) 
        &=2\rmi[a^{-}_{\alpha}(q)-a^{+}_{\alpha}(q)]\tilde{\Pi}_{\mu\nu}(q)-\frac{1}{2(q \cdot n)}\Xi_{\mu\nu\alpha}\,,
\end{align}
where $a^{\mu}_{\pm}(q) \coloneqq \rmi c^{(\pm)\dagger}_{\rm R}\partial_{\mu}^{q}c^{(\pm)}_{\rm R}$ is the Berry connection and $\Xi_{\mu\rho\alpha}\coloneqq -\tfrac{\rmi}{2}\hat{q}_{\perp(\mu}S_{\rho)\alpha}+\hat{q}_{\perp[\mu}\Delta_{\rho]\alpha} $.
According to these results, eq.~\eqref{eq:W2} can be written as
\begin{equation}
    \begin{split}
        &W^{\text{R}}_{\mu\nu\rho\sigma}(x,q)\\
        &=64\pi^2 G\delta(q^2)\text{sgn}(q\cdot n)\times\\ &\quad\left[\qty(\text{Re}\qty[\Pi_{\mu\nu\rho\sigma}]-\frac{\hbar}{2}\text{Im}\qty[\Sigma_{\mu\nu\rho\sigma\alpha}]\partial^{\alpha})+\rmi\qty(\text{Im}\qty[\Pi_{\mu\nu\rho\sigma}]+\frac{\hbar}{2}\text{Re}\qty[\Sigma_{\mu\nu\rho\sigma\alpha}]\partial^{\alpha})\right]\check{f}_{\text{R}}(x,q)\\
        &\simeq 64\pi^2 G \delta(q^2)\text{sgn}(q\cdot n)\qty[\Pi_{\mu\nu\rho\sigma}-\frac{\rmi\hbar}{8(q \cdot n)}(\Xi_{(\mu\rho\alpha}\tilde{\Pi}_{\nu)\sigma}+\Xi_{(\nu\sigma\alpha}\tilde{\Pi}_{\mu)\rho})\partial^{\alpha}]f_{\text{R}}(x,q)\\
        &=\calS^{\text{R}}_{\mu\nu\rho\sigma}+\rmi\calA^{\text{R}}_{\mu\nu\rho\sigma},
    \end{split}
\end{equation}
where we have introduced a frame-dependent distribution function by absorbing the Berry connection,
\begin{equation}
    f_{\text{R}}(x,q)\coloneqq \qty(1+2\hbar [a^{\alpha}_{+}(q)-a^{\alpha}_{-}(q)]\partial_{\alpha})\check{f}_{\text{R}}(x,q).
\end{equation}
Finally, using eqs.~\eqref{eq:SR} and \eqref{eq:AR}, the lesser propagator for left-handed gravitons can be obtained by replacing the spin-tensor $S_{\mu\nu}$ with $-S_{\mu\nu}$.

\section{Transport equation}\label{app:transport_equation}
The graviton transport equation can be derived from the same strategy as in refs.~\cite{Liu:2018xip, Mameda:2022ojk}.
We impose the Wigner translation and the linearized Einstein equation, $\bar{\nabla}^2 h_{\mu\nu} =0 $, in the Lorenz and TT gauges under the weak gravity, on the two-point correlation function:
\begin{equation}\label{eq:two_point_cor}
    \begin{split}
        G_{\alpha\beta|\mu\nu\rho\sigma}=\int _{y}\rme^{-\rmi \frac{q\cdot y}{\hbar}}\left\langle h^{+}_{\rho\sigma}\qty(\frac{1}{2}D_{(\alpha}-\partial^{y}_{(\alpha})\qty(\frac{1}{2}D_{\beta)}-\partial^{y}_{\beta)})h_{\mu\nu}^{-}\right\rangle,\\
    \end{split}
\end{equation}
where $h^{\pm}_{\mu\nu}\coloneqq h_{\mu\nu}(x,\pm \frac{y}{2})$.
Here we introduce the following shorthand notations for the operators:
\begin{align}
 \calG_\mu 
  &= -\frac{\rmi y^\nu}{2\hbar}\sum_{n=0}^\infty\frac{\bigl[\calC(y\cdot D)\bigr]^n}{(n+2)!} 
  \, G_{\mu\nu} \,, \\
  \calH_\mu 
  &= -\frac{\rmi y^\nu}{\hbar}\sum_{n=0}^\infty\frac{\bigl[\calC(y\cdot D)\bigr]^n}{(n+1)!} 
  \, G_{\mu\nu} \,, \\
  G_{\mu\nu} 
  &=-\rmi\hbar [D_\mu,D_\nu] \,,
\end{align}
where $\calC(Y) Z = [Y,Z]$.
Using these notations, eq.~\eqref{eq:two_point_cor} can be written as
\begin{equation}
    \begin{split}\label{eq:G1}
        G_{\alpha\beta|\mu\nu\rho\sigma}
        &=\qty(\frac{1}{2}D_{(\alpha}-\frac{\rmi q_{(\alpha}}{\hbar})\qty(\frac{1}{2}D_{\beta)}-\frac{\rmi q_{\beta)}}{\hbar})W_{\mu\nu\rho\sigma}\\
        &\qquad+\qty(\frac{1}{2}D_{(\alpha}-\frac{\rmi q_{(\alpha}}{\hbar})\int_{y}\rme^{-\rmi \frac{q\cdot y}{\hbar}}\left\langle\calG_{\beta)}h_{\rho\sigma}^{+}h_{\mu\nu}^{-} \right\rangle\\
        &\qquad\qquad+\int_{y}\rme^{-\rmi \frac{q\cdot y}{\hbar}}\left\langle -\qty[\qty(\frac{1}{2}D_{(\alpha}-\partial^{y}_{(\alpha})\calG_{\beta)}h^{+}_{\rho\sigma}]h^{-}_{\mu\nu} \right\rangle.
    \end{split}
\end{equation}
Alternatively, performing integration by parts yields
\begin{equation}
    \begin{split}\label{eq:G2}
        &G_{\alpha\beta|\mu\nu\rho\sigma}\\
        &=\int_{y} \rme^{-\rmi \frac{q\cdot y}{\hbar}}\left\langle h^{+}_{\rho\sigma}\qty[\qty(\frac{3}{2}D_{(\alpha}-\partial^{y}_{(\alpha})\calG_{\beta)}-\qty(D_{(\alpha}\calH_{\beta)}+\calH_{(\alpha}D_{\beta)}+\calH_{(\alpha}\calH_{\beta)})]h^{-}_{\mu\nu} \right\rangle\\
        &\qquad+\int_{y}\rme^{-\rmi \frac{q\cdot y}{\hbar}}\left\langle h^{+}_{\rho\sigma}\rme^{-\frac{y\cdot D}{2}}\nabla_{(\alpha}\nabla_{\beta)}h_{\mu\nu}(x)\right\rangle.
    \end{split}
\end{equation}
Equations~\eqref{eq:G1} and ~\eqref{eq:G2} lead to
\begin{equation}
    \begin{split}
        \bar{g}^{\alpha\beta}\qty(q_{(\alpha}+\frac{\rmi\hbar}{2}D_{(\alpha})\qty(q_{\beta)}+\frac{\rmi\hbar}{2}D_{\beta)})W_{\mu\nu\rho\sigma}=\mathcal{O}(\hbar^2).
    \end{split}
\end{equation}

\bibliographystyle{JHEP}
\bibliography{ref}
\end{document}